\documentstyle[sprocl]{article}

\def\be{\beta}
\def\ga{\gamma}
\def\de{\delta}
\def\ep{\epsilon}

\def\ka{\kappa}
\def\la{\lambda}

\def\si{\sigma}

\def\ph{\phi}

\def\ch{\chi}
\def\ps{\psi}

\def\Ga{\Gamma}
\def\De{\Delta}

\def\cl{{\cal L}}

\def\prt{\partial}
\def\vev#1{\langle {#1}\rangle}

\def\fr#1#2{{{#1} \over {#2}}}
\def\frac#1#2{{\textstyle{{#1}\over {#2}}}}
\def\half{{\textstyle{1\over 2}}}
\def\lsim{\mathrel{\rlap{\lower4pt\hbox{\hskip1pt$\sim$}}
    \raise1pt\hbox{$<$}}}
\def\gsim{\mathrel{\rlap{\lower4pt\hbox{\hskip1pt$\sim$}}
    \raise1pt\hbox{$>$}}}
\def\sqr#1#2{{\vcenter{\vbox{\hrule height.#2pt
         \hbox{\vrule width.#2pt height#1pt \kern#1pt
         \vrule width.#2pt}
         \hrule height.#2pt}}}}

\def\lrprtmu{\stackrel{\leftrightarrow}{\partial_\mu}}
\def\lrprtnu{\stackrel{\leftrightarrow}{\partial^\nu}}

\def\Im{\hbox{Im}\,}

\newcommand{\beq}{\begin{equation}}
\newcommand{\eeq}{\end{equation}}
\newcommand{\bea}{\begin{eqnarray}}
\newcommand{\eea}{\end{eqnarray}}
\newcommand{\rf}[1]{(\ref{#1})}

\begin{document}

\begin{flushright}
IUHET 397\\
October 1998
\end{flushright}
\vglue 0.5 cm

\title{THE STATUS OF CPT\footnote
{Talk presented at WEIN 98, 
Santa Fe, New Mexico, June 1998} 
}

\author{V.\ ALAN KOSTELECK\'Y}

\address{Physics Department, Indiana University\\
Bloomington, IN 47405, U.S.A.\\
Email: kostelec@indiana.edu} 

\maketitle\abstracts{ 
A short review is given of some theoretical approaches 
to CPT violation.
A potentially realistic possibility is that small
apparent breaking of CPT and Lorentz symmetry 
could arise at the level of the standard model
from spontaneous symmetry breaking in an underlying theory.
Some experimental constraints are described.
}

\section{Introduction}

Among the observed symmetries of nature 
are CPT and Lorentz invariance.
The discrete transformation CPT is the product of 
charge conjugation C, 
parity reflection P, 
and time reversal T,
while the Lorentz transformations 
include rotations and boosts.
These symmetries are connected via the CPT theorem,
which under mild assumptions states that CPT  
is an exact symmetry of local Lorentz-covariant field theories 
of point particles.\cite{cpt,sachs}

Both CPT and Lorentz invariance have been tested
to a high degree of precision
and in a variety of experiments.
For example,
the sharpest figure of merit for CPT tests 
quoted by the Particle Data Group\cite{pdg}
involves the kaon particle-antiparticle mass difference,
which has been bounded by experiments at Fermilab and CERN
to\cite{kexpt}
\beq
r_K \equiv \fr{|m_K - m_{\overline K}|}{m_K}
\lsim 10^{-18}
\quad .
\label{a}
\eeq
At present,
CPT is the sole combination of C, P, T
observed as an exact symmetry of nature at the fundamental level.

Since the CPT theorem holds generally 
for relativistic particle theories
and since there exist high-precision experimental tests,
the observation of CPT violaton 
would represent a powerful signal for 
unconventional physics.
It is therefore of interest to examine
possible theoretical mechanisms 
through which CPT might be broken. 

In the next part of this talk,
I briefly review some approaches that have been taken
in the literature to address the possibility of CPT violation.
It turns out that most suggestions either have physical features
that seem unlikely to be realized in nature 
and/or involve radical revisions
of conventional quantum field theory. 
However,
there exists at least one promising possibility,
based on spontaneous breaking of CPT
and Lorentz symmetry,\cite{ks,kp}
that appears to be compatible with experimental constraints
and with established quantum field theory.
Its formulation and experimental implications
are described in later parts of this talk.

\section{Approaches to CPT Violation}

Perhaps the simplest approach to CPT violation
is a purely phenomenological one,
avoiding the issue of developing a
microscopic theory allowing CPT breaking.
This can be implemented for a given 
experimental situation
by introducing a parametrization 
of the observable quantities
that allows for the possibility of CPT violation.
The method has the advantages
that it can be relatively straightforward 
in principle
and that it is to some degree independent 
of possible origins of the CPT-violating effects.
Among the disadvantages are the impossibility 
of relating the bounds obtained to those 
from other experiments
and the absence of predictive power.

A well-established example of the phenomenological approach
to CPT violation can be found in the literature
on kaon oscillations.\cite{leewu}
The physical eigenstates $K_S$ and $K_L$ can be expressed
as linear combinations of the strong-interaction eigenstates
$K^0$ and $\overline{K^0}$.
Two complex parameters denoted $\ep_K$ and $\de_K$ 
appear in these combinations.
Both parametrize CP violation,
but $\ep_K$ governs T violation 
while $\de_K$ governs CPT violation.
The standard model of particle physics
has a mechanism for T violation,
and so $\ep_K$ is at least in principle 
a calculable and nonzero quantity.
However, 
the standard model preserves CPT
and so predicts that $\de_K$ is identically zero. 
Allowing for a nonzero value of $\de_K$
is from this viewpoint a purely phenomenological choice.
It has no grounds in a microscopic theory
and $\de_K$ is therefore not a calculable quantity.
Moreover,
it cannot be linked to other phenomenological parameters 
for CPT tests in different experiments.

A more interesting (and harder) 
approach from the theoretical perspective
is to construct an explicit microscopic theory
for CPT violation.
Any such effort must somehow avoid 
one or more of the assumptions of the CPT theorem.

An immediate possibility is to construct a theory
that directly violates one of the major axioms
leading to the CPT theorem.
For example,
nonlocal field theories might be considered.
Examples of these have been provided by Carruthers,\cite{car}
who studied a class of models that are Lorentz covariant
and involve conventional quantization
but for which the imposition of self-conjugacy 
and half-integer isospin suffices to produce nonlocal field operators.
In these models
CPT is broken,
and the violation of the CPT theorem
can be traced to the nonlocality of the operators.
No multiplets of this type are known in nature.

A more subtle possibility is to consider models that
violate one of the technical requirements of the CPT theorem
that otherwise might appear relatively unimportant.
For example,
one assumption of the CPT theorem is that
the fields lie in finite-dimensional representations
of the Lorentz group.
Oksak and Todorov\cite{ot}
have given examples of models
that involve infinite-spin multiplets
but are Lorentz covariant.
Despite the Lorentz covariance,
these models break CPT.
The seeming failure of the CPT theorem
is a consequence of the appearance of
infinite-dimensional Lorentz representations
needed to describe the infinite-spin fields.
Again, 
no such multiplets have been identified in nature. 

Another approach is to consider models beyond 
the framework of particle field theory.
For example,
string (M) theories are qualitatively different 
from particle theories because they involve extended objects
and it is unclear \it a priori \rm 
whether the CPT theorem applies.
Although certain solutions to a subset of string models 
are known to be CPT invariant,\cite{pas}
it has been shown\cite{kp} that 
in some string theories CPT violation may occur 
through a mechanism based on spontaneous breaking
of Lorentz symmetry.\cite{ks}
This mechanism can be understood within
conventional quantum field theory.\cite{kp2,cksm}
It may lead to observable effects 
at the level of the standard
model,\cite{kp2}$^{\!-\,}$\cite{bkr2}
as is described in later sections below.
At the level of the standard model,
the appearance of CPT violation is compatible
with the CPT theorem because it is accompanied
by breaking of the Lorentz symmetry.

A more radical suggestion has been made by Hawking\cite{sh}
that quantum mechanics might be violated
by effects from quantum gravity
and that CPT violation might be among the consequences.
It is unclear how to incorporate effects of this type
in the context of a conventional field theory
such as the standard model,
which relies on the usual structure of quantum mechanics.
An extension of this idea has been 
suggested in the context of string theory.\cite{je}
It would produce a signature in the kaon system
requiring at least six\cite{bf} 
phenomenological parameters other than $\de_K$.

In searching for attractive possibilities for CPT violation,
the ideal would be a microscopic theory 
valid at a fundamental level
that also provides a quantitative connection 
to experiment in the framework of the standard model.
This would then allow 
the calculation of phenomenological parameters,
direct comparisons between experiments,
and perhaps the prediction of signals.
Some progress towards the development of
such a theory has been made
in the context of the idea
of spontaneous CPT symmetry breaking,
which is described in the next section.

\section{Spontaneous CPT Violation}

Even if the underlying theory of nature 
has a Lorentz- and CPT-covariant action,
apparent violations of these symmetries 
could result from spontanteous symmetry 
breaking.\cite{ks,kp}
To my knowledge,
there are at present no theoretical problems
that would appear to exclude the possibility of  
small spontaneous Lorentz breaking,
so this could represent a relatively attractive way to 
violate CPT and Lorentz invariance.
Moreover,
spontaneous Lorentz breaking of some type must be
a property of any Lorentz-covariant higher-dimensional 
theory that purports to underly nature
because only four macroscopic dimensions are observed. 

In general,
spontaneous breaking is merely a feature of the solutions
and leaves unchanged the symmetry of the underlying dynamics,
and so it hides a symmetry rather than directly breaking it.
Many of the desirable features of 
a Lorentz-covariant theory would therefore be expected
to remain intact under spontaneous Lorentz breaking,
as distinct from other types of Lorentz breaking
that are likely to be inconsistent 
with desirable theoretical properties.
For example,
microcausality can be explicitly verified
in certain simple models arising
from spontaneous Lorentz breaking.\cite{cksm}
Indeed,
the physics of a particle in a Lorentz-breaking vacuum
is in some respects similar to that of
a particle moving inside a biaxial crystal.
The behavior of the latter 
is not typically rotation or boost Lorentz covariant.
Rather than being a fundamental problem,
this is merely indicative of the presence 
of the background crystal fields,
and properties such as causality remain unaffected.

Spontaneous breaking of Lorentz symmetry could occur in 
a theory with Lorentz-covariant dynamics
that contains certain types of interaction among Lorentz-tensor fields.
If such interactions destabilize the naive vacuum 
and produce nontrivial expectation values,
then the presence in the true vacuum of a small 
Lorentz-tensor expectation means that
Lorentz symmetry is spontaneously broken.\cite{ks}
This mechanism may occur in some string theories
because suitable interactions are known to appear,
unlike the case of
conventional four-dimensional renormalizable gauge theories
such as the standard model.
If any of the tensor expectation values
involves a field with an odd number of spacetime indices,
CPT is spontaneously broken too.\cite{kp} 
If any components of the expectation values lie 
along the four macroscopic spacetime dimensions,
apparent violations of Lorentz symmetry 
and possibly also CPT could emerge
at the level of the standard model.\cite{kp2}

For the (unrealistic) case of the open bosonic string,
the mechanism of spontaneous Lorentz breaking
can be investigated using string field theory.
The action and the corresponding equations of motion
can be obtained analytically  
for particle fields below some fixed level number $N$.
Deriving and comparing solutions for different $N$
permits the identification 
of solutions that persist as $N$ is increased.\cite{ks}
In some cases this procedure has been performed 
to a depth of over 20,000 terms in 
the static potential.\cite{kp}
Among the solutions found are ones spontaneously breaking
Lorentz invariance that remain stable as $N$ increases.

If Lorentz symmetry is regarded as global,
then its spontaneous breaking 
would entail the appearance of massless modes
in accordance with the Nambu-Goldstone theorem.
When gravity is included,
Lorentz invariance is promoted to a local symmetry. 
In conventional gauge theories,
the promotion of a global spontaneously broken symmetry 
to a local one is associated with the Higgs mechanism,
by which the massless modes disappear
and the vector-boson propagator is modified 
to include a mass term.
However, 
there is no analogous effect in gravity:~\cite{ks}
when local Lorentz symmetry is spontaneously broken,
the graviton propagator is affected
but the dependence of the connection on
derivatives of the metric rather than the metric itself
ensures that no graviton mass is generated.

\section{Standard-Model Extension and QED Limit}

Since there is at present no compelling evidence
for either Lorentz or CPT violation,
any effects from spontaneous breaking
must be suppressed at the level of the
minimal SU(3)$\times$SU(2)$\times$U(1) standard model.
If the relevant dimensionless suppression factor 
is determined by the ratio of the scale of the
standard model to the scale of an underlying fundamental theory
at the Planck mass,
only a few observable effects of Lorentz or CPT violation 
are likely to exist.
These effects should be derivable from an extension
of the standard model that 
is obtained as the low-energy limit 
of the fundamental theory.\cite{kp2}

As an example,
consider the following class of possible additional terms 
in the fermionic sector of
the low-energy limit of the underlying theory:
\beq
\cl \sim \fr {\la} {M^k} 
\vev{T}\cdot\overline{\ps}\Ga(i\prt )^k\ch
+ {\textstyle h.c.}
\quad .
\label{aa}
\eeq
Terms of this type could arise,
for instance,
from the coupling between one or more bosonic tensor fields
and fermion bilinears
when the tensors acquire an expectation value $\vev{T}$.
In the above expression,
the bilinear in the fermion fields $\ps$, $\ch$ 
contains a gamma-matrix structure $\Ga$
and the coupling involves $k$ spacetime derivatives $i\prt$,
which together would produce apparent Lorentz and CPT violation 
in the low-energy theory.
The coupling constant in this example
is a combination of a dimensionless coupling $\la$
and a suitable power of a large scale $M$
associated with the fundamental theory,
such as the Planck or compatification scale.

An analysis of this type can be used to incorporate
the effects of spontaneous Lorentz and CPT breaking  
at the level of the standard model.
The procedure is to add to the lagrangian 
all possible extra terms that apparently break these symmetries
and that could arise from spontaneous symmetry breaking
in a more fundamental theory.

By restricting attention to the subset of 
allowed hermitian terms that preserve both
SU(3) $\times$ SU(2)$ \times$ U(1) gauge invariance
and power-counting renormalizability,
a general Lorentz-violating extension 
of the standard model
that includes both CPT-even and CPT-odd terms 
has been constructed.\cite{cksm}
This theory appears at present to be the sole candidate
for a consistent extension of the standard model
based on a microscopic theory of Lorentz violation.
By construction, 
it must be the low-energy limit
of any underlying theory (not necessarily string theory) 
that contains spontaneous Lorentz and CPT violation
and that reduces to the standard model
in the limiting case of exact Lorentz invariance.

As might be anticipated from the discussion 
of spontaneous symmetry breaking in the previous section, 
this standard-model extension displays several attractive features
despite the apparent violation of Lorentz and CPT symmetry.\cite{cksm}
Since Lorentz covariance is a property of the underlying theory,
properties like microcausality and positivity of the energy
are to be expected. 
Also,
since the standard-model extension is based
on otherwise conventional field theory,
the usual quantization methods are unaffected.
Provided the vacuum tensor expectation values 
arising from the spontaneous breaking 
are independent of spacetime position,
i.e., disregarding possible solitonic solutions,
energy and momentum are conserved. 
Covariance under rotations or boosts of the observer's inertial frame
(observer Lorentz transformations)
remains a feature of the theory.
The apparent Lorentz violations appear
only when (localized) fields are rotated or boosted
(particle Lorentz transformations)
relative to the vacuum tensor expectation values.
Moreover,
although not evident \it a priori, \rm
it turns out that the usual  
gauge symmetry breaking to the electromagnetic U(1) occurs.

Many of the high-precision experiments 
sensitive to CPT and Lorentz violation 
are associated with quantum electrodynamics (QED).
It is therefore useful to extract from the
standard-model extensions various limiting cases
that represent generalizations of the usual versions of QED. 
Modifications to QED 
from Lorentz- and CPT-breaking effects 
can appear in both the photon and fermion sectors.\cite{cksm}

As an example,
consider the limiting case of the standard-model extension
that reduces in the absence of Lorentz breaking
to the normal quantum field theory 
of photons, electrons, and positrons.
The usual lagrangian is:
\beq
\cl^{\rm QED} =
\overline{\ps} \ga^\mu (\half i \lrprtmu - q A_\mu ) \ps 
- m \overline{\ps} \ps 
- \frac 1 4 F_{\mu\nu}F^{\mu\nu}
\quad .
\label{aaa}
\eeq
Extra terms that break Lorentz invariance
appear in both the photon and fermion sectors,
and they can be CPT even or CPT odd.
The CPT-violating terms are:
\beq
\cl^{\rm CPT}_{e} =
- a_{\mu} \overline{\ps} \ga^{\mu} \ps 
- b_{\mu} \overline{\ps} \ga_5 \ga^{\mu} \ps \quad ,
$$
$$
\cl^{\rm CPT}_{\ga} =
\half (k_{AF})^\ka \ep_{\ka\la\mu\nu} A^\la F^{\mu\nu}
\quad .
\label{bbb}
\eeq
The CPT-even terms are:
\beq
\cl^{\rm Lorentz}_{e} = 
c_{\mu\nu} \overline{\ps} \ga^{\mu} 
(\half i \lrprtnu - q A^\nu ) \ps 
+ d_{\mu\nu} \overline{\ps} \ga_5 \ga^\mu 
(\half i \lrprtnu - q A^\nu ) \ps 
- \half H_{\mu\nu} \overline{\ps} \si^{\mu\nu} \ps 
$$
$$
\cl^{\rm Lorentz}_{\ga} =
-\frac 1 4 (k_F)_{\ka\la\mu\nu} F^{\ka\la}F^{\mu\nu}
\quad .
\label{ccc}
\eeq
All these terms violate particle Lorentz covariance,
although observer Lorentz covariance is maintained.
The conventions and notation used in these equations
are discussed in the literature,\cite{cksm}
along with various other issues.
The coefficients of the extra terms above
behave as Lorentz- and CPT-violating couplings,
and in accordance with the discussion at the 
beginning of this section
they are expected to be minuscule. 
Note that field redefinitions can be used 
to demonstrate that not all the
components are physically observable.
For example, 
coefficients of the type $a_\mu$
can only be detected directly in flavor-changing experiments
and so are unobservable at leading order in any situation 
where only electrons, positrons, and photons are involved.

\section{Experiments}

Present-day experiments  
seeking evidence for the Lorentz-violating couplings
in the standard-model extension
face the difficult task of overcoming
a suppression factor likely to be 
about 17 orders of magnitude,
comparable to the ratio of the 
standard-model and Planck scales.
Most experiments would lack the necessary precision 
to detect possible signals,
but a few exceptionally sensitive tests
can already place interesting bounds on
some of the coupling coefficients.

The standard-model extension 
described in the previous section
provides a quantitative basis
within which to analyze and compare different experiments
on CPT and Lorentz symmetry,
and in some situations it can suggest possibilities
for observable signals.
In this context,
several existing and planned experimental tests 
have been studied.
They include observations of 
neutral-meson oscillations,\cite{kexpt,kp,cksm}$^{\!-\,}$\cite{ak}
comparative tests of QED 
in Penning traps,\cite{bkr,pennexpts}
spectroscopy of hydrogen and antihydrogen,\cite{bkr2,antih}
measurements of cosmological birefringence,\cite{cksm}
and observations of the baryon asymmetry.\cite{bckp}
In the remaining parts of this talk,
a short summary of a subset of these investigations
is given.
Other work along these lines and
currently underway includes a study\cite{kla}
of limits attainable in
clock-comparison experiments.\cite{cc}

\subsection{Neutral-Meson Oscillations}

Several neutral-meson systems exhibit
or are expected to exhibit flavor oscillations,
including $K$, $D$, $B_d$, and $B_s$.
The time evolution of a neutral-meson state 
is controlled by a two-by-two effective hamiltonian
in the meson-antimeson state space.
Denoting the neutral meson by $P$,
this non-hermitian hamiltonian contains complex parameters
$\ep_P$ and $\de_P$
that govern (indirect) CP violation.
For the $K$ system,
these are the same phenomenological quantities 
already mentioned in the section of this talk 
about approaches to CPT violation.
The parameter $\ep_P$ measures 
T violation with CPT invariance,
while $\de_P$ measures
CPT violation with T invariance.
Experiments observing $P$-meson oscillations
can constrain the magnitude of $\de_P$ 
and hence place limits on possible CPT breaking.

As mentioned before,
$\de_P$ is necessarily zero in the context of the
usual standard model,
which preserves CPT.
However,
in the context of the standard-model extension
an expression for $\de_P$ can be derived.\cite{ak}
Remarkably,
at leading order this expression depends only 
on one particular kind of extra coupling
in the standard-model extension,
of the form
$- a^q_{\mu} \overline{q} \ga^\mu q$.
Here, 
$q$ represents one of the valence quark fields
in the $P$ meson,
and the quantity $a^q_{\mu}$ is spacetime constant 
but depends on the quark flavor $q$.

The presence of Lorentz breaking
means that the expression for $\de_P$
varies with the boost and orientation of the $P$ meson.\cite{ak}
If the $P$-meson four-velocity is given as 
$\be^\mu \equiv \ga(1,\vec\be)$ 
in the frame in which the quantities $a^q_{\mu}$ are specified, 
then
$\de_P$ is given at leading order in all coupling coefficients by
\beq
\de_P \approx i \sin\hat\ph \exp(i\hat\ph) 
\ga(\De a_0 - \vec \be \cdot \De \vec a) /\De m
\quad .
\label{e}
\eeq
Subscripts $P$ have been omitted on the right-hand side
for simplicity.
In this expression,
$\De a_\mu \equiv a_\mu^{q_2} - a_\mu^{q_1}$,
where $q_1$ and $q_2$ are the valence-quark flavors 
for the $P$ meson.
Also,
$\hat\ph\equiv \tan^{-1}(2\De m/\De\ga)$,
where $\De m$ and $\De \ga$
are the mass and decay-rate differences
between the $P$-meson eigenstates,
respectively.

An immediate implication of this result is
that tests of CPT and Lorentz symmetry with neutral mesons
are independent at leading order of other types of tests
mentioned in this talk.
The point is that $\de_P$ is sensitive \it only \rm
to $a^q_{\mu}$,
and moreover this is due to flavor-changing effects.
No other tests mentioned here involve flavor changes,
and so,
as mentioned at the end of the previous section,
it can be shown that none can observe 
effects from nonzero values of $a^q_{\mu}$.

The result \rf{e} also has direct implications for
experiments with neutral mesons.
It predicts\cite{kp2}
that the real and imaginary 
parts of $\de_P$ are proportional and that 
the magnitude of $\de_P$ may be different for different $P$ 
due to the flavor dependence 
of the coefficients $a_\mu^q$.
It is even conceivable that the heavier neutral mesons
such as $D$ or $B_d$ exhibit much larger CPT-violating
effects if,
for instance,
the coefficients $a_\mu^q$ behave like conventional
Yukawa couplings and grow with mass.

A more striking prediction is that 
signals for Lorentz and CPT violation 
in neutral-meson experiments would depend on
the boost magnitude and orientation
of the mesons involved,
which implies several effects.\cite{ak}
One is that experiments with otherwise comparable 
statistical sensitivity to CPT effects
may in fact have inequivalent CPT reach.
This might happen if the mesons involved have
very different momentum spectra
or if they are well collimated as opposed to
having a $4\pi$ distribution.

The tightest experimental constraints on CPT violation
presently in the literature 
come from observations of the $K$ system.\cite{kexpt}
The possibility of CPT violation in the
heavier neutral-meson systems has received 
relatively little experimental attention,
although two collaborations\cite{bexpt}
at CERN have performed analyses 
to study the possibility\cite{ckv}
that existing data could suffice to constrain CPT violation.
A measurement 
$\Im\de_{B_d} = -0.020 \pm 0.016 \pm 0.006$
has been published by the OPAL collaboration,
while a preliminary result of 
$\Im\de_{B_d} = -0.011 \pm 0.017 \pm 0.005$
has been given by the DELPHI collaboration.
There are additional theoretical and experimental
studies in progress.

\subsection{QED Experiments}

An ingenious type of high-precision experiment 
is based on the idea of trapping individual particles
for extended time periods
so that accurate measurements of properties can be made.
Comparisons of results for particles and antiparticles 
then provide useful CPT tests.
It can be shown that these experiments
are sensitive to effects in the 
fermion sector of the QED extension.\cite{bkr}

Comparative measurements of particle and antiparticle 
anomaly and cyclotron frequencies 
have been obtained using Penning traps.\cite{pennexpts}
In the context of the QED extension,
there are both direct signals and
effects arising from diurnal variations
in a comoving Earth-laboratory frame. 
\cite{bkr}
Appropriate figures of merit for the various signals
have been defined
and the attainable experimental sensitivity estimated.

As one example,
experiments comparing the anomalous magnetic moments
of electrons and positrons could generate a sharp bound
on the spatial components of the coefficient $b_\mu$
in the laboratory frame.
A minor change in experimental procedure
would permit a bound of order $10^{-20}$ 
on the associated figure of merit
to be obtained with existing technology,
and indeed data from a suitable experiment
are now being analyzed.\cite{rm} 
Similar experiments for protons and antiprotons might be envisaged.

Another possibility is to compare cyclotron frequencies 
of various particles and antiparticles.
An ingenious experiment 
comparing the cyclotron frequencies of $H^-$ ions
and antiprotons in the same trap
has been performed by Gabrielse
and coworkers.\cite{gk}
In the context of the standard-model extension,
the leading-order effects in this experiment
provide a test of Lorentz violation
with an associated figure of merit 
bounded at $4\times 10^{-25}$.

A somewhat different class of tests can be performed with trapped 
hydrogen and antihydrogen.\cite{antih}
The idea is to obtain high-precision spectroscopic data
for the two systems,
which can then be compared to provide CPT tests.
Within the context of the standard-model and QED extensions,
an investigation of the possible experimental signals
involving 1S-2S and hyperfine transitions 
has been completed.\cite{bkr2}
It turns out that specific transitions 
for magnetically trapped hydrogen and antihydrogen
are directly sensitive to CPT- and Lorentz-violating couplings,
without any suppression factors
associated with the fine-structure constant.
Moreover, 
certain experimental tests could in principle provide 
a theoretically clean signal for particular
types of coupling that break Lorentz and CPT symmetry.

A variety of constraints 
can be placed on the photon sector of the QED extension
from theoretical considerations
and from terrestrial, astrophysical,
and cosmological experiments on electromagnetic waves.
On the theoretical front,
a consistency constraint may arise 
on the pure-photon term appearing in Eq.\ \rf{bbb} 
because it can provide negative contributions
to the energy.\cite{cfj}
This contrasts with the CPT-even term  
appearing in the following equation, 
which maintains a positive conserved energy,
and suggests the coefficient
$(k_{AF})^\ka$ should be zero.\cite{cksm}

A theoretical treatment of the extended Maxwell equations 
including CPT- and Lorentz-breaking effects
shows that,
as usual, 
the solutions involve 
two independent propagating degrees of freedom.
However,
unlike the conventional propagation of 
electromagnetic waves in vacuum,
the dispersion relations of the two modes differ
and so the vacuum is birefringent.
The effects of the Lorentz and CPT violation 
on an electromagnetic wave traveling in the vacuum
are closely analogous to those exhibited by 
an electromagnetic wave in conventional electrodynamics 
that is passing through 
an optically anisotropic and gyrotropic transparent crystal 
with spatial dispersion of the axes.\cite{cksm}

On the experimental front,
the tightest constraints emerge 
from the observed absence of birefringence 
on cosmological distance scales.
For the pure-photon term in Eq.\ \rf{bbb},
this absence translates into a bound on the
components of the CPT-odd coefficient $(k_{AF})_\mu$ 
of the order of $\lsim 10^{-42}$ GeV.\cite{cfj,hpk}
A disputed claim\cite{nr,misc}
for a nonzero effect at the level of 
$|\vec k_{AF}|\sim 10^{-41}$ GeV has been made.

For the pure-photon term in Eq.\ (5),
the single rotation-invariant irreducible component of 
the CPT-even coefficient $(k_F)_{\ka\la\mu\nu}$ 
is constrained to $\lsim 10^{-23}$ 
by the existence of cosmic rays\cite{cg}
and other tests.
All other irreducible components
of $(k_F)_{\ka\la\mu\nu}$ 
break rotation invariance.
At present,
no bounds from cosmological birefringence 
have been placed on these components,
but in principle they could be constrained experimentally 
with known techniques.\cite{cksm}
An estimate suggests
the dimensionless coefficient $(k_F)_{\ka\la\mu\nu}$ 
could be bounded at the level of about $10^{-27}$.

Evidently,
the zero value of $(k_{AF})_\mu$ 
needed to avoid negative-energy contributions
is compatible with 
the tight experimental constraints obtained.
However,
since no symmetry protects a zero 
tree-level value of $(k_{AF})_\mu$,
one might expect 
$(k_{AF})_\mu$ to be shifted away from zero
by radiative corrections 
involving CPT-violating couplings
in the fermion sector.
Remarkably,
it turns out\cite{cksm}
that the one-loop radiative corrections are finite,
which means
a tree-level CPT-odd term is unnecessary
for one-loop renormalizability.
Higher loops may exhibit similar effects.
Note that there is no similar mechanism
for the CPT-even pure-photon term,
for which calculations have explicitly demonstrated\cite{cksm}
the existence of divergent radiative corrections
at the one-loop level
and which would leave open the possibility of
detecting a nonzero effect in cosmological birefringence.
The feasibility of setting to zero an otherwise allowed
CPT-odd pure-photon term represents a nontrivial
consistency check on the standard-model extension.

As a final remark about possible observable CPT effects,
it has been shown\cite{bckp}
that under appropriate conditions
the observed baryon asymmetry could be produced
in thermal equilibrium
as a result of the existence of CPT-violating bilinear terms
of the general form in Eq.\ \rf{aa}.
At grand-unified scales,
a relatively large baryon asymmetry could be generated
that could ultimately be diluted
to the observed value through sphaleron or other effects.
This would represent an alternative to conventional baryogenesis,
for which nonequilibrium processes
and C- and CP-breaking interactions are required.\cite{ads}

\section*{Acknowledgments}
I thank Orfeu Bertolami, Robert Bluhm, Don Colladay, 
Rob Potting, Neil Russell, Stuart Samuel, 
and Rick Van Kooten for collaborations.
This work was supported in part
by the United States Department of Energy 
under grant number DE-FG02-91ER40661.

\end{document}